\documentclass[oneside]{IEEEtran}  

\usepackage[cmex10]{amsmath}  
\usepackage{amssymb}
\usepackage{cite}
\usepackage{url}
\usepackage{graphicx}
\usepackage{booktabs}
\usepackage{array}
\usepackage[hidelinks]{hyperref}

\usepackage[caption=false,font=footnotesize]{subfig}

\renewcommand{\tablename}{Table}
\renewcommand{\figurename}{Figure}

\newcommand{\safegraphic}[3]{%
  \IfFileExists{#1}{%
    \includegraphics[width=#2]{#1}%
  }{%
    \fbox{%
      \parbox[c][#3][c]{#2}{%
        \centering
        \textbf{Figure Placeholder}\\[2mm]
        \texttt{\detokenize{#1}}\\[1mm]
        Replace this box with the final artwork.%
      }%
    }%
  }%
}

\begin{document}


\title{\textsc{Low-Altitude Agentic Networks: From UAV Connectivity to Workflow-Driven Capability Orchestration}}

\author{Baolin~Chong, Fuhui~Zhou, Qihui~Wu, and Ran~Tu%
\thanks{The authors are with Nanjing University of Aeronautics and Astronautics.}%
}

\maketitle


\section*{\large{Abstract}}

Low-altitude platforms are evolving from connected sensing nodes into
autonomous agents capable of bounded task interpretation, skill invocation,
and local decision making. Increasing the intelligence of individual
unmanned aerial vehicles (UAVs), however, does not resolve the organization
problem that arises when mission workflows depend on capabilities distributed
across multiple platforms and provider feasibility changes with network,
resource, and mobility conditions. We propose low-altitude agentic networking
as a task-level networking paradigm and develop a reference architecture that
separates workflow logic from executor identity. Within a validated
domain-specific workflow space, each ready operation is translated into a
capability requirement and dynamically bound to an agent that remains
feasible under current conditions. Task contracts and structured evidence
govern workflow execution, while routine provider and link changes are handled
through workflow-preserving local adaptation before high-level repair is
invoked. A trace-calibrated discrete-event case study identifies a bounded
operating region: conventional organization remains effective for simple
workflows under stable conditions, whereas agentic networking becomes
beneficial when workflow dependencies are tightly coupled, capability
feasibility changes during execution, and viable local recovery paths remain
available.


\IEEEpeerreviewmaketitle


\section*{\large{I. Introduction}}

Beyond their traditional roles as remotely controlled sensing,
communication, and execution platforms, unmanned aerial vehicles (UAVs) are
increasingly equipped with multimodal perception, task-oriented reasoning,
and agent runtimes \cite{dong2026aerial,nguyen2026agentic,zhang2026multiuav}. Within a bounded mission domain, an agent-enabled UAV can
interpret task instructions, invoke specialized skills, and contribute
task-level decisions. This evolution strengthens local autonomy, but it does
not automatically organize the intelligence and capabilities distributed
across multiple platforms.

Complex low-altitude missions often require different UAVs to perform
complementary operations and depend on the timely delivery of intermediate
results to downstream executors. Meanwhile, the suitability of a capability
provider may change during execution. A UAV may possess the specialized skill
required by an operation yet become unable to continue serving the active task
because its view is obstructed, its workload is excessive, or its remaining
energy is insufficient. Capability must therefore describe not only what an
agent can perform in principle, but also whether that capability can actually
be delivered under the current network, resource, and mobility conditions \cite{zeng2016wireless,mozaffari2019tutorial}.

Centralized coordination can effectively organize UAV operations when timely
global state is available, while distributed task allocation can reconsider
the assignment of known tasks without relying on a permanent controller \cite{choi2009cbba}.
Agent runtimes can further improve local task interpretation and skill
invocation \cite{yao2023react,zhang2026multiuav}. However, these paradigms do not necessarily maintain the evolving
relation among workflow requirements, capability feasibility, and current
providers as part of the network's runtime state. The unresolved problem is
how to form task-specific collaboration relations and preserve workflow
execution as provider availability and communication conditions change.

To address this problem, we propose low-altitude agentic networking as a new
task-level networking paradigm for organizing distributed UAV intelligence
and capabilities around mission workflows. The proposed paradigm separates
task logic from executor identity. A mission is instantiated within a
domain-specific and validated workflow space \cite{qiao2025worfbench,wang2025awm,li2026graphflow}, and each ready workflow
operation is translated into a capability requirement that is dynamically
bound to an agent that remains feasible under current conditions \cite{chen2025ioa,guo2025discovery}. Routine
provider and link changes are handled through workflow-preserving local
adaptation. High-level reasoning is invoked for repair only when the active
workflow no longer has a feasible continuation. We further develop a
reference architecture for this paradigm and use a trace-calibrated
discrete-event case study to identify the operating conditions under which
its benefits outweigh the additional coordination overhead.

The remainder of this article is organized as follows. Sections~II--IV
establish the conceptual boundary, reference architecture, and networking
foundations of the proposed paradigm. Section~V presents the case study,
Section~VI discusses open research problems, and Section~VII concludes the
article.


\begin{figure*}[!t]
  \centering
  \includegraphics[width=0.7\textwidth]{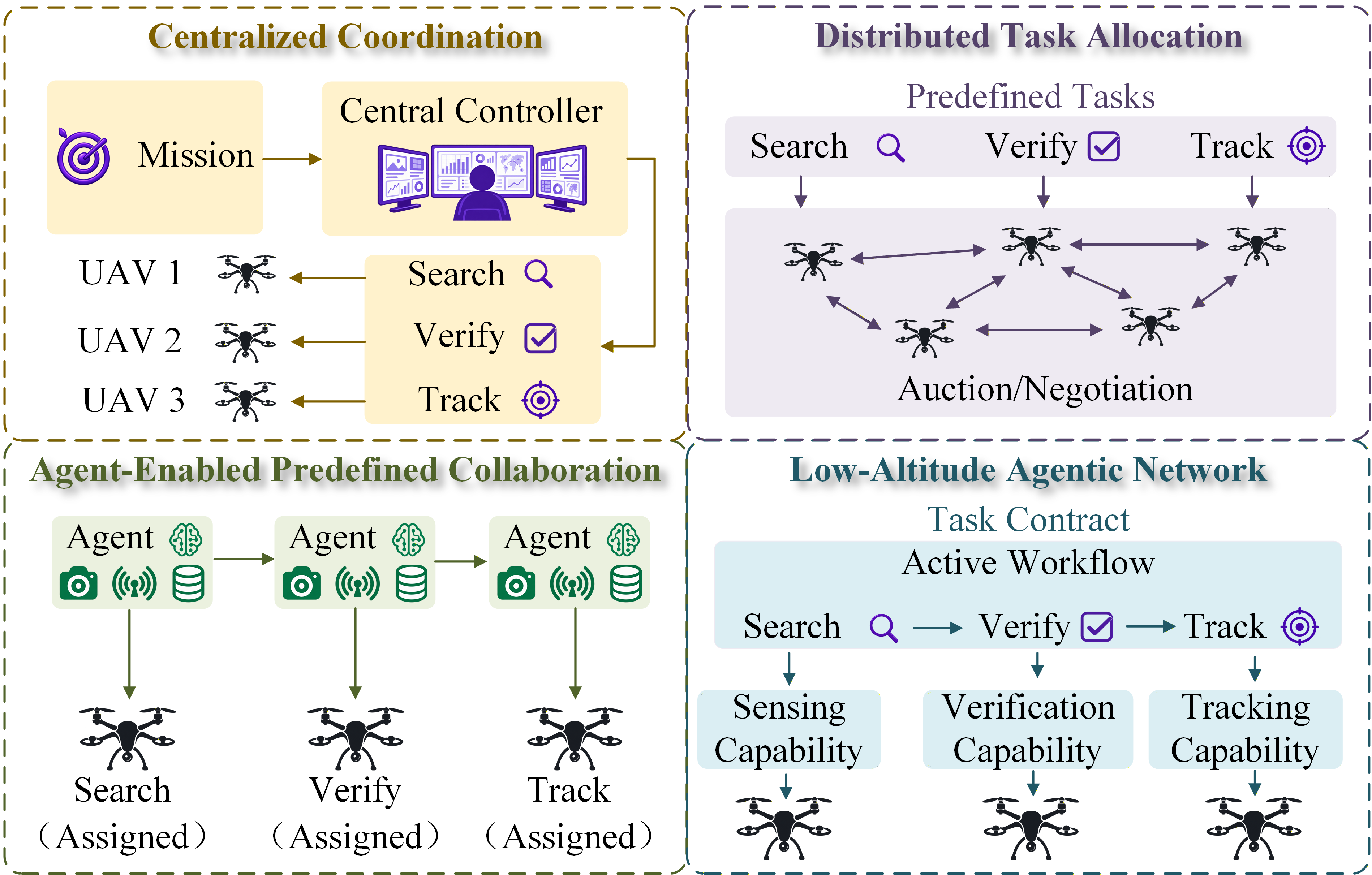}
  \caption{Four organizational paradigms for low-altitude UAV systems,
  distinguished by how task relations are formed and maintained.}
  \label{fig:paradigm_evolution}
\end{figure*}

\section*{\large{II. From UAV Networks to Low-Altitude Agentic Networks}}

Fig.~\ref{fig:paradigm_evolution} contrasts four representative
organizational paradigms according to how task relations are formed and
maintained. Their defining characteristics are summarized below.

\subsection*{A. Centralized UAV Networks}

In a centralized UAV network, a controller interprets the mission and assigns
its operations to UAVs. This organization is efficient when the environment
is stable and global state remains current. When an executor becomes
unavailable, however, the affected task relation must be re-established
through the controller. Recovery timeliness therefore depends on the global
control loop even when a suitable replacement exists nearby.

\subsection*{B. Distributed Task Allocation}

Distributed methods move executor selection from a controller to negotiation
among UAVs, allowing newly arrived tasks or failed assignments to be handled
locally \cite{choi2009cbba}. These methods typically operate on tasks and dependencies defined in
advance. They decentralize executor selection, but generally operate on task
operations and dependencies that have already been defined.

\subsection*{C. Agent-Enabled UAVs with Predefined Collaboration}

An agent runtime allows a UAV to interpret bounded instructions and invoke
specialized skills, thereby improving node-level autonomy \cite{dong2026aerial,nguyen2026agentic,zhang2026multiuav}. When roles and
collaboration partners are configured in advance, however, the mission still
follows predefined relations. An agent may adjust its own action or invoke a
designated backup, but it does not necessarily search the network for the most
suitable provider of an emerging capability requirement. Node intelligence
determines how an assigned operation is performed, whereas network-level
intelligence is also needed to determine how collaboration relations are
formed.

\subsection*{D. Low-Altitude Agentic Networks}

We propose low-altitude agentic networking as a task-level networking paradigm
in which collaboration relations are formed and maintained around mission
workflows. Recent work has used the same term for an optical oceanic
communication and sensing architecture \cite{mao2026optical}; here it denotes
a workflow-driven task-level organization paradigm. Rather than permanently associating workflow operations with
predefined UAVs, the system selects a task-appropriate subgraph from a
validated workflow space. Each ready operation is translated into a capability
requirement and dynamically bound to an agent that remains feasible under
current conditions. Task logic is therefore separated from executor identity,
allowing the same operation to be served by different agents as execution
conditions evolve.

Owning a required skill does not necessarily make an agent a feasible
provider. The agent must also be able to execute the operation and deliver its
result under the current task, network, resource, and mobility conditions.
When a provider becomes infeasible, the system first preserves the active
workflow by adapting its communication path or binding the affected operation
to another agent. The workflow itself is repaired only when no feasible
realization can continue to satisfy the task objective.


\begin{figure*}[!t]
  \centering
  \includegraphics[width=0.7\textwidth]{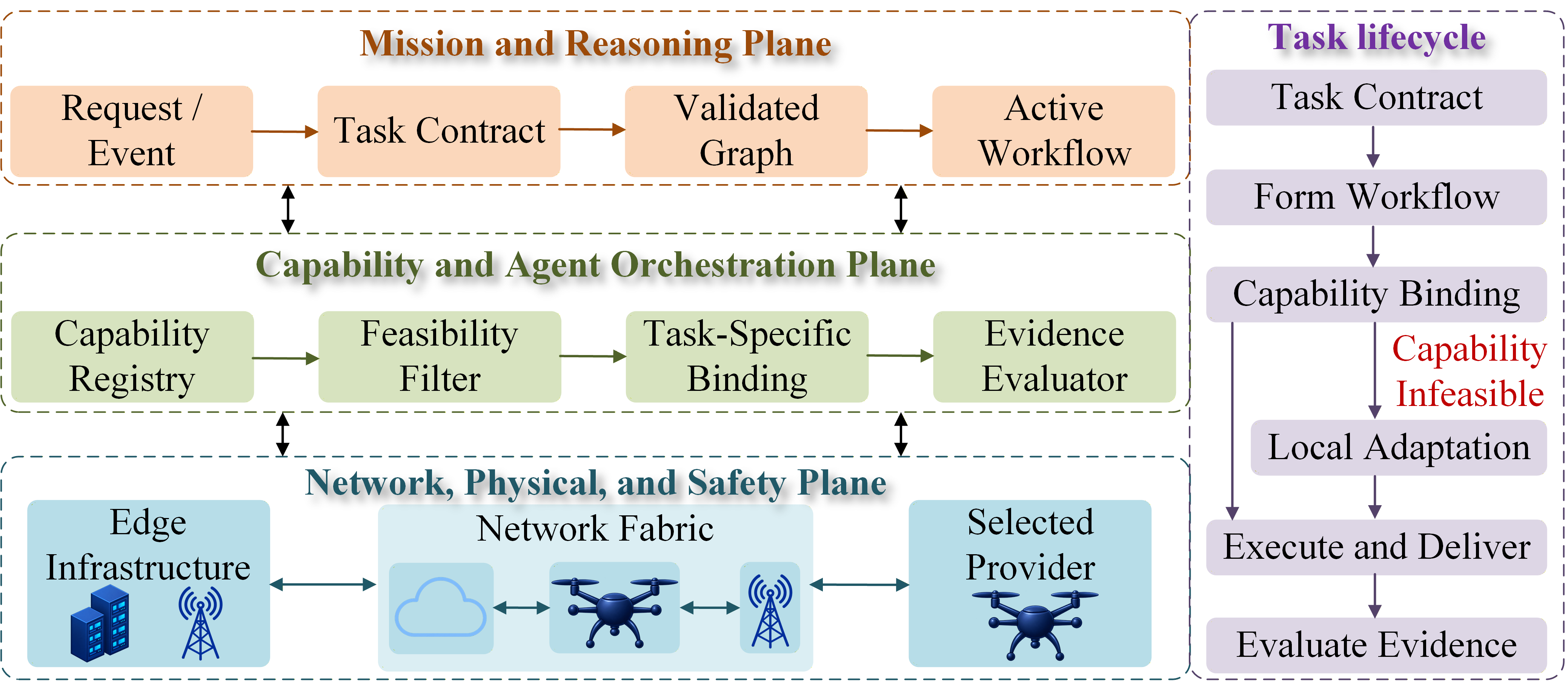}
  \caption{Reference architecture and task lifecycle of the proposed
  low-altitude agentic network, from task-contract formation to
  evidence-driven adaptation.}
  \label{fig:architecture_lifecycle}
\end{figure*}

\section*{\large{III. Reference Architecture and Task Lifecycle}}

This section presents the reference architecture of the proposed
low-altitude agentic network and explains how a mission is transformed into a
bounded workflow, bound to feasible agents, and adapted during execution.

\subsection*{A. Three-Plane Architecture and Hierarchical Intelligence}

Fig.~\ref{fig:architecture_lifecycle} illustrates three interacting planes.
The mission and reasoning plane interprets an external mission or a
task-relevant event, constructs a task contract, and activates an appropriate
workflow. The capability and agent orchestration plane translates each ready
workflow operation into a capability requirement and maintains its binding to
a feasible agent. The network, physical, and safety plane supports
communication and platform actions while enforcing deterministic operational
constraints. Execution results return as evidence and update the workflow
state.

An agent-enabled UAV should not be identified with a single language model.
Its agent core maintains local task context, interprets bounded instructions,
and invokes specialized skills. Ground or edge large language models (LLMs)
handle mission-level interpretation and uncommon workflow repair, whereas an
onboard small language model (SLM), or a lightweight deterministic runtime on
a constrained platform, supports routine decisions and local skill
invocation. High-frequency perception, navigation, tracking, communication
control, and flight control remain with specialized models and deterministic
controllers \cite{dong2026aerial,nguyen2026agentic}.

\subsection*{B. Task Origination, Contracts, and Evidence}

A high-level mission typically enters through a ground or edge mission agent,
while new subtasks may be derived during execution from local evidence,
delegation, or failure events. A UAV that detects a possible target may request
verification, while a tracking agent whose communication path degrades may
request relay support. The architecture therefore combines a controlled
mission entry with distributed task derivation, without requiring every UAV to
act as a complete mission planner.

A task contract provides a machine-readable interface between mission intent,
workflow execution, and network support. It specifies the intended outcome,
required inputs and evidence, and the temporal, reliability, resource, and
safety constraints that must be satisfied. When a subtask is delegated, the
receiving agent obtains the relevant contract and evidence rather than the
complete natural-language mission history.

Operational state and execution evidence serve different purposes. State
describes the current conditions used to select a provider, such as position,
workload, energy, and connectivity. Evidence records the result produced by an
executed operation, such as a detected target, a verification confidence, or a
tracking update. Structured evidence is evaluated directly by the workflow
runtime, while ambiguous or conflicting evidence may be escalated for
model-assisted interpretation.

\begin{figure*}[!t]
  \centering
  \includegraphics[width=0.7\textwidth]{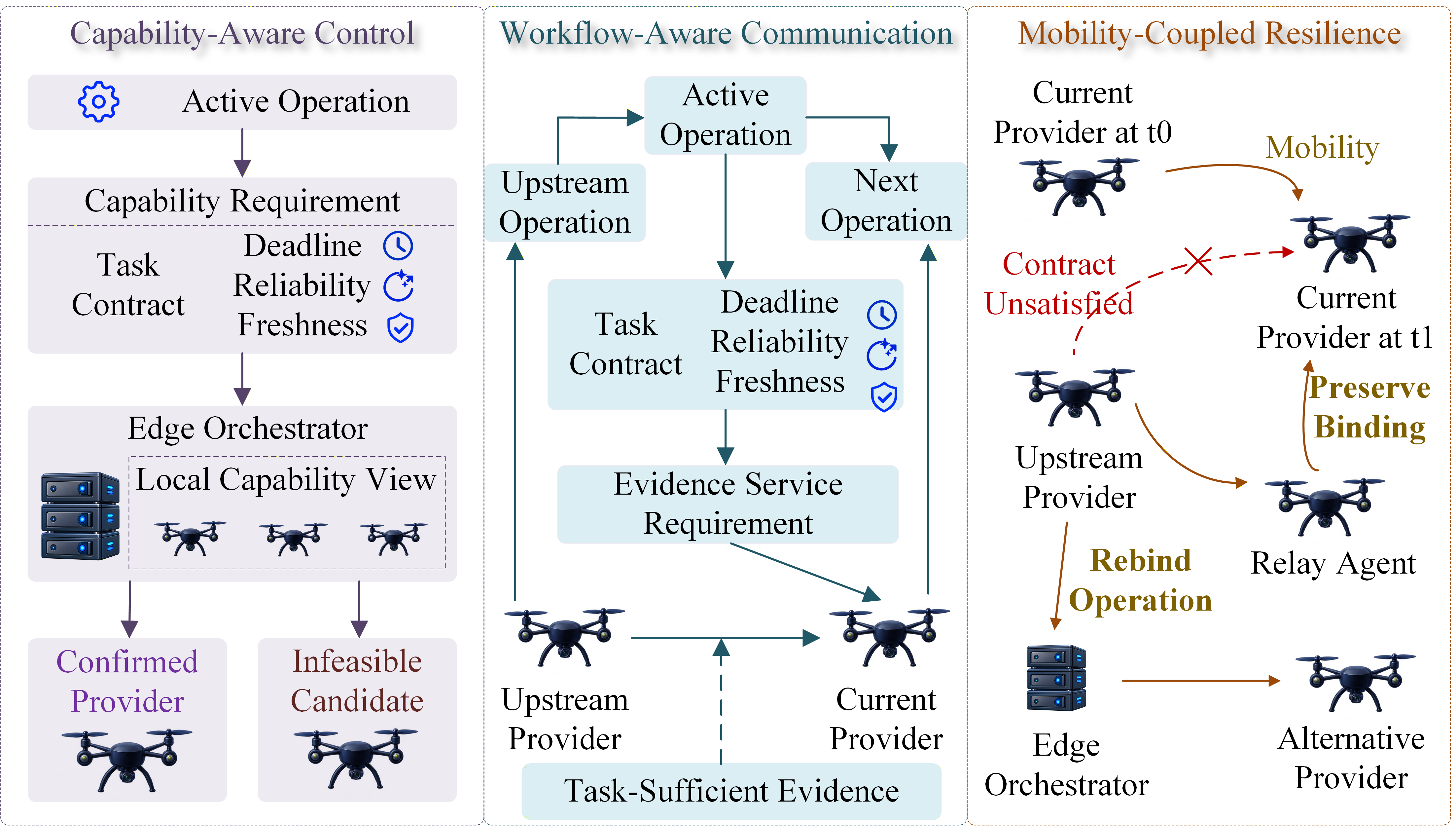}
  \caption{Networking support for agentic collaboration through
  capability-aware control, workflow-aware communication, and
  mobility-coupled resilience.}
  \label{fig:networking_support}
\end{figure*}

\subsection*{C. Graph-Assisted Workflow Formation and Capability Binding}

Workflow formation takes place within a domain-specific and validated workflow
space. This space may be constructed offline from expert-defined mission
templates, successful execution traces, and model-generated candidate paths
\cite{wang2025awm,zhuge2024gptswarm}.
Its operations and dependencies are admitted only after their interfaces,
preconditions, evidence requirements, structural validity, and safety
constraints have been checked.

For an incoming task contract, the reasoning plane retrieves and instantiates
a task-appropriate subgraph rather than asking an LLM to generate an
unrestricted workflow from scratch \cite{qiao2025worfbench,li2026graphflow}. Evidence determines which validated branch
is activated during execution. A confident verification result may activate
tracking, whereas an uncertain result may activate independent verification.
An LLM proposes a constrained repair only when the active subgraph has no
valid continuation that can satisfy the mission contract.

Each ready workflow operation identifies a required capability rather than a
particular UAV. The capability registry first identifies agents that declare
the relevant skill and then determines which candidates remain feasible under
the current task, network, resource, mobility, and safety conditions.
Infeasible candidates are removed, and the remaining providers are ranked
according to expected service quality and completion cost. Binding an
operation only when it becomes ready allows provider selection to reflect the
latest system state.

This separation keeps task logic independent of executor identity. A workflow
operation can be reassigned to another UAV without discarding completed
operations or valid evidence, allowing the workflow to remain stable while its
physical realization changes.

\subsection*{D. End-to-End Execution and Workflow-Preserving Adaptation}

Consider a search--verify--track mission. The mission contract activates a
search operation and binds it to a feasible sensing provider. Candidate-target
evidence then makes verification ready. If the verification evidence satisfies
the required confidence, tracking is activated; otherwise, the workflow
follows its validated second-verification branch. Since each operation is
bound only when it becomes ready, the selected provider reflects the latest
target, agent, and network state.

During tracking, the provider must continue to satisfy evidence-freshness,
delivery-reliability, and delay requirements. A UAV may retain its tracking
skill while becoming infeasible because its observation or communication path
no longer satisfies the active contract. The runtime first attempts to
preserve the current workflow by restoring the path, introducing a relay,
repositioning a platform, or rebinding the affected operation. These actions
change the execution relation while preserving the mission objective and
accumulated evidence. High-level workflow repair is invoked only when no
feasible local response can maintain the active task contract.

\begin{figure*}[!t]
    \centering

    \subfloat[Mission success rate.]{
        \includegraphics[width=0.32\textwidth]{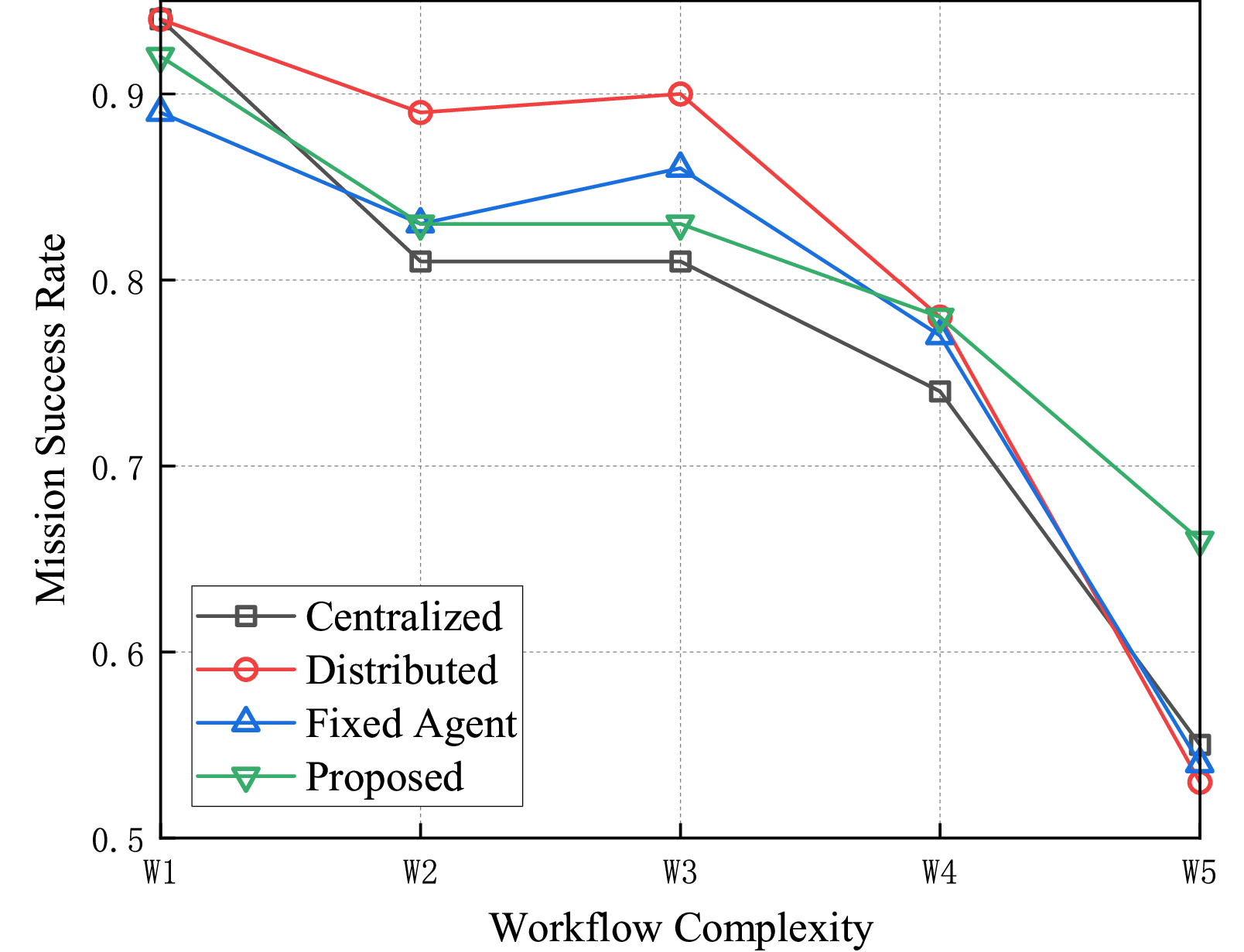}
        \label{fig:success_rate}
    }
    \subfloat[Restricted mean completion time.]{
        \includegraphics[width=0.32\textwidth]{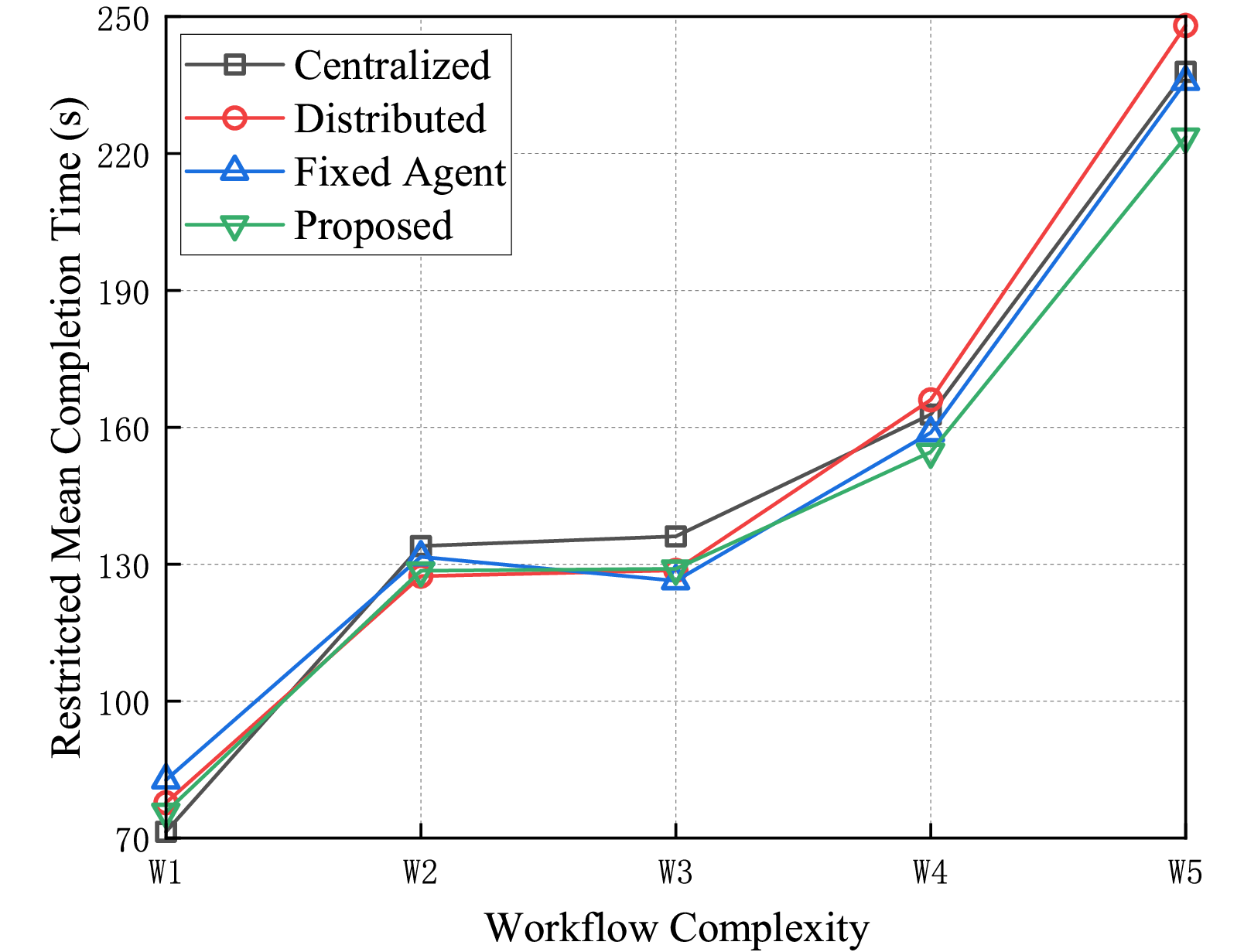}
        \label{fig:completion_time}
    }
    \subfloat[Success-rate advantage over the best baseline.]{
        \includegraphics[width=0.32\textwidth]{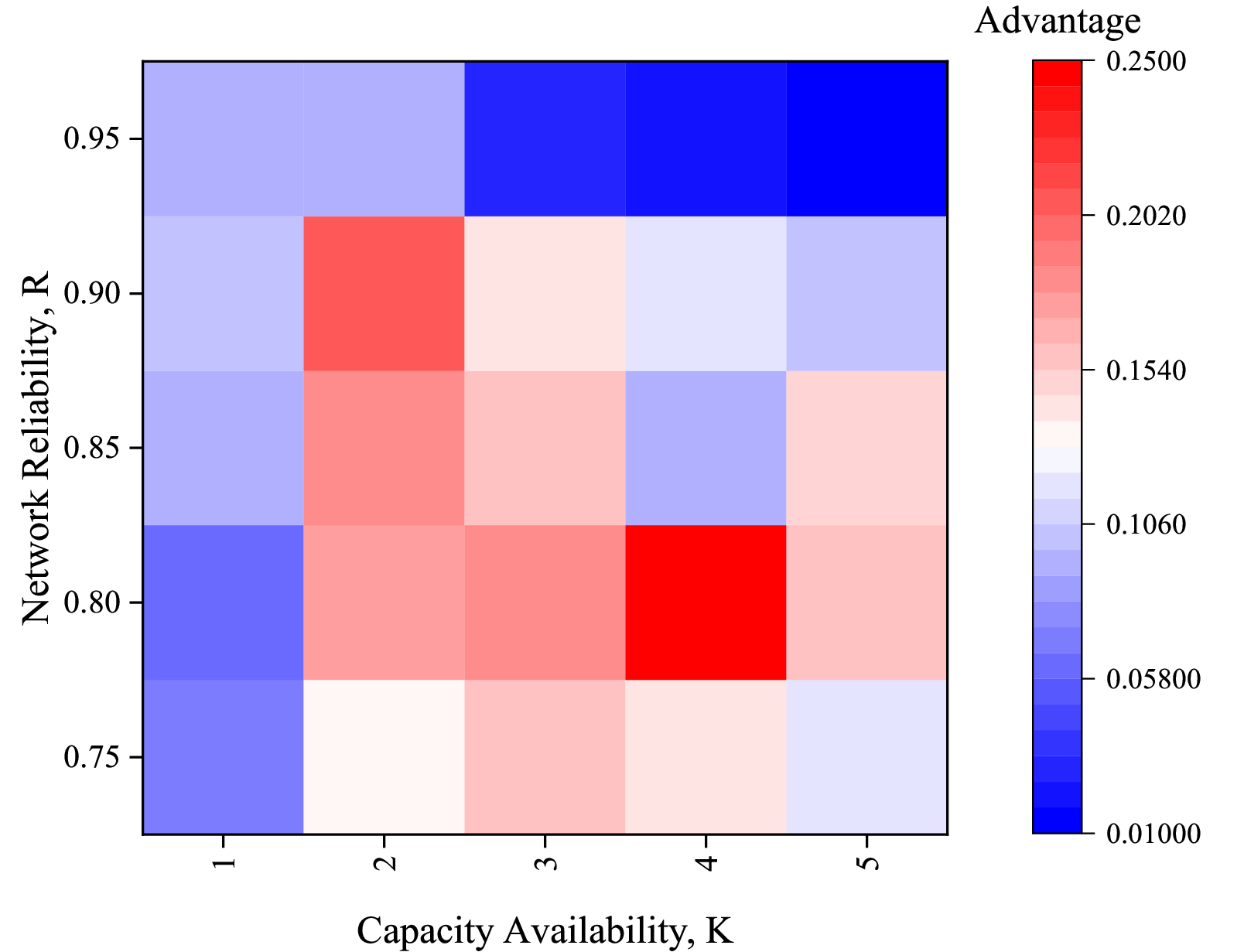}
        \label{fig:advantage_heatmap}
    }

    \caption{Results of Experiments 1 and 2. Outlined cells in (c) satisfy
    the predefined statistical and efficiency criteria for the strict
    advantage region.}
    \label{fig:evaluation_results}
\end{figure*}

\section*{\large{IV. Networking Foundations for Agentic Collaboration}}

Section~III explained how workflow operations are translated into capability
requirements and bound to feasible agents. The network must support this
process before, during, and after binding. It exposes sufficiently fresh
information about candidate providers, carries task contracts and execution
evidence between the selected agents, and updates the execution relation when
network, resource, or mobility conditions change. The resulting communication
structure therefore follows the active workflow rather than a permanent
collaboration graph.

\subsection*{A. Capability-Aware Network Control}

Capability-aware control extends conventional node identity and reachability
with the service that an agent can currently provide. A capability record
combines a relatively stable description of the supported skill with dynamic
information about the provider's position, workload, energy, availability,
network reachability, and the validity interval of that state. This allows the
orchestration plane to distinguish declared capability from task-specific
feasibility. An agent may own the required skill yet remain unsuitable if it
cannot complete the operation or deliver its evidence under the active task
contract \cite{guo2025discovery}.

Maintaining a perfectly synchronized capability view is neither realistic nor
necessary. Mobility, workload, and resource consumption may invalidate a
record, whereas reporting every state change would consume resources needed by
the mission. A practical control plane can combine periodic advertisements
with event-triggered updates and on-demand confirmation before a critical
binding. Discovery may begin from a local view and expand only when nearby
information is insufficient. The objective is a task-sufficient capability
view that is fresh enough for the next binding or recovery decision.

\subsection*{B. Workflow-Aware Communication and Orchestration}

A dependency between workflow operations creates a temporary communication
relation between their current providers. When verification activates
tracking, the verification evidence must reach the agent selected for the
tracking operation. If tracking is later rebound, the communication endpoints
must change accordingly. The workflow therefore determines which agents need
to communicate, while the task contract specifies the delay, reliability, and
freshness requirements of their exchange \cite{chen2025ioa,zhuge2024gptswarm}.

The transmitted representation should match the needs of the downstream
operation. A specialized skill may require an original observation or an
intermediate feature, whereas a workflow transition may require only a compact
structured result. The network should carry a representation that is
sufficient for the downstream task without assuming that every exchange can
be reduced to a semantic message. Evidence on the active workflow path should
also receive priority over background capability updates or noncritical agent
traffic. Provider selection, routing, and communication-resource allocation
should therefore be coordinated around timely workflow completion rather than
optimized independently.

\subsection*{C. Mobility-Coupled Resilience}

Mobility is both a source of disruption and a resource for recovery. Movement
changes sensing geometry, obstruction, and communication quality, so a
provider that was feasible when selected may later become unable to satisfy
the active task contract. The agent may retain the required skill and remain
operational, yet no longer provide a feasible capability because its evidence
cannot reach the next workflow stage with the required delay, reliability, or
freshness \cite{zeng2016wireless,mozaffari2019tutorial}.

Recovery should preserve as much of the current workflow and accumulated
evidence as possible. An alternative route or relay may restore the existing
provider; otherwise, the affected operation can be rebound to another feasible
agent. Repositioning may be used when physical movement can recover sensing or
communication feasibility without violating safety constraints or disrupting
another critical task. These actions differ in delay, energy consumption, and
task impact, so the network should select the least disruptive feasible
response. High-level workflow repair is invoked only when no local networking
or physical action can continue to satisfy the task contract.

This ordering establishes a fast adaptation path within the networking and
orchestration planes. Routine provider and connectivity changes are absorbed
through routing, relaying, rebinding, or repositioning, while model-assisted
reasoning is reserved for situations in which the validated workflow has no
feasible continuation. The network can thereby preserve workflow execution
without turning every transient change into a new mission-planning problem.


\begin{table}[t]
    \centering
    \caption{Mobility-induced workflow recovery over 50 paired seeds.
    Continuity denotes the fraction of the required Tracking interval that
    satisfies the Tracking Contract.}
    \label{tab:mobility_recovery}

    \scriptsize
    \renewcommand{\arraystretch}{1.15}
    \setlength{\tabcolsep}{3pt}

    \begin{tabular}{
        >{\raggedright\arraybackslash}m{1.15cm}
        >{\centering\arraybackslash}m{1.15cm}
        >{\centering\arraybackslash}m{1.55cm}
        >{\centering\arraybackslash}m{1.55cm}
        >{\centering\arraybackslash}m{1.10cm}
    }
        \toprule
        \textbf{Policy} &
        \textbf{Success Rate} &
        \textbf{Median Recovery (s)} &
        \textbf{Mean Disruption (s)} &
        \textbf{Continuity} \\
        \midrule

        Centralized &
        \textbf{1.00} &
        9.10 &
        \underline{9.57} &
        \underline{0.787} \\

        Distributed &
        \underline{0.92} &
        9.43 &
        13.94 &
        0.729 \\

        Fixed Agent &
        0.82 &
        \underline{8.17} &
        18.45 &
        0.678 \\

        Proposed &
        \textbf{1.00} &
        \textbf{4.89} &
        \textbf{5.10} &
        \textbf{0.887} \\

        \bottomrule
    \end{tabular}
\end{table}

\section*{\large{V. Case Study}}

This section evaluates dynamic agentic organization from three perspectives:
its benefit under increasing workflow collaboration complexity, its operating
region under different capability and network conditions, and its recovery
performance under low-altitude mobility.

\begin{figure*}[!t]
  \centering
  \includegraphics[width=0.7\textwidth]{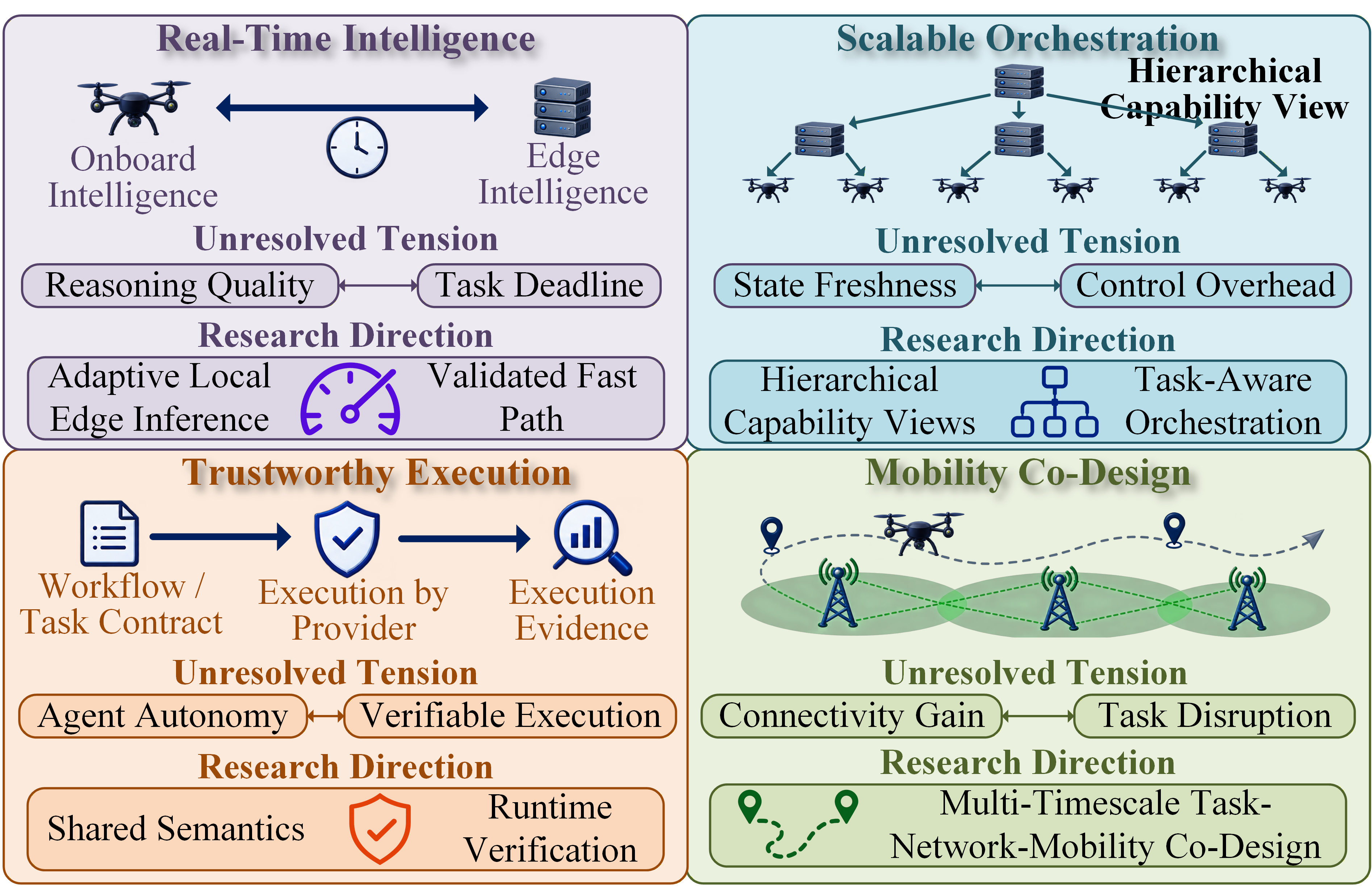}
  \caption{Open research problems in real-time intelligence, scalable
  orchestration, trustworthy execution, and mobility co-design.}
  \label{fig:open_research}
\end{figure*}

\subsection*{A. Simulation Setup}

The discrete-event simulation framework integrates a Workflow Runtime,
Contract Manager, Capability Registry, and Network Emulator. Large-scale
episodes were executed on a CPU. Qwen2.5-1.5B-Instruct and
Qwen2.5-7B-Instruct were profiled in BF16 on an NVIDIA A40 \cite{qwen2024qwen25}, and the measured
traces calibrated model-scale-dependent reasoning delays rather than a
specific onboard computing platform. Experiments 1 and 2 used 16
heterogeneous logical agents in a
$1{,}000~\mathrm{m}\times1{,}000~\mathrm{m}$ area with a 300-s episode limit.
Experiment 3 used eight UAVs and one ground/edge coordination node with a
130-s limit. The Tracking Contract required an end-to-end evidence-delivery
reliability of at least 0.90 and a delay of at most 0.20~s, with recovery
triggered after three consecutive invalid updates.

Four organization policies were compared. \textbf{Centralized} periodically
refreshed global state and performed global replanning. \textbf{Distributed}
used a bounded auction with at most three communication rounds, with bids
reflecting expected completion time, workload, and communication cost
\cite{choi2009cbba}.
\textbf{Fixed Agent} followed predefined Primary, Backup, and Relay
assignments. \textbf{Proposed} dynamically bound feasible providers and
selected among Reroute, Rebind, Relay, and Reposition. Paired episode
manifests provided all policies with identical initial states, skill outcomes,
target states, disruption events, and model traces.

Experiment 1 increased workflow collaboration complexity at $K=3$ and
$R=0.95$. $W_1$ contained Search--Verify; $W_2$ added continuous Tracking;
$W_3$ added evidence-dependent re-verification; $W_4$ introduced recoverable
link degradation; and $W_5$ added two-provider Search, Evidence Fusion, and
repeated disruptions. Experiment 2 fixed $W_4$ and scanned
$K\in\{1,2,3,4,5\}$ and
$R\in\{0.75,0.80,0.85,0.90,0.95\}$, where $K$ is the number of
operationally eligible providers per critical capability before communication
path feasibility is considered and $R$ is the pre-recovery probability that a
task-relevant transfer meets its deadline. Experiment 3 introduced
mobility-induced building obstruction that made the initial tracker's path
contract-infeasible.

Mission success rate (SR) was the primary metric. Experiment 1 also reported
restricted mean completion time (RMCT), computed with the 300-s episode cap.
In Experiment 2, a cell entered the strict advantage region when
$\Delta SR\geq0.05$, the lower bound of the paired-bootstrap 95\% confidence
interval of $\Delta SR$ was positive, control overhead was within $2\times$
that of the best baseline, and completion time was within $1.2\times$ that of
the best baseline. Control overhead counted orchestration messages and their
serialized bytes.

\subsection*{B. Evaluation Results}

Figs.~\ref{fig:evaluation_results}(a) and
\ref{fig:evaluation_results}(b) show a complexity-dependent benefit. For
$W_1$--$W_3$, the policies remain close because these workflows require
limited reorganization. At $W_4$, Proposed matches the highest baseline SR of
0.78 and achieves the shortest RMCT. The advantage becomes pronounced at
$W_5$: Proposed reaches an SR of 0.66, compared with 0.55 for the best
baseline, while reducing RMCT from 235.77~s to 223.64~s. Dynamic organization
is therefore most valuable when coupled evidence dependencies and recurring
disruptions allow local binding failures to propagate through the workflow.

Fig.~\ref{fig:evaluation_results}(c) further identifies where this advantage
appears. At $R=0.95$, stable links limit the benefit of rebinding, and
$\Delta SR$ remains between 0.01 and 0.09. Larger gains occur mainly at lower
reliability with sufficient provider redundancy, reaching 0.18 in the strict
advantage region. Some cells with large raw differences remain unoutlined
because they fail at least one statistical or efficiency criterion. The result
shows that the benefit depends jointly on network variation and the
availability of feasible alternatives rather than on increasingly adverse
conditions alone.

Table~\ref{tab:mobility_recovery} reports mobility-induced recovery. Proposed
and Centralized both achieve an SR of 1.00. Relative to Centralized, Proposed
reduces median recovery time from 9.10~s to 4.89~s and mean disruption time
from 9.57~s to 5.10~s, while increasing continuity from 0.787 to 0.887. These
results show that contract-triggered local adaptation restores the workflow
before the remote global control loop completes replanning.

\section*{\large{VI. Open Research Problems}}

The mechanisms developed in Sections~III and IV expose four connected research
problems. Fig.~\ref{fig:open_research} links each architectural mechanism to
its unresolved tension and a representative direction, keeping the discussion
grounded in the proposed system rather than in generic concerns about
artificial intelligence.

\subsection*{A. Real-Time Intelligence}

Low-altitude missions require decisions within a limited time, while a UAV
cannot continuously support complex reasoning. Ground-side inference offers a
stronger model but adds communication delay and may become unreachable after a
link failure. An onboard model responds quickly but may not handle an event
outside the active workflow. A late high-quality decision can be useless,
whereas an insufficiently verified local decision may propagate an error to
later operations. A promising direction is adaptive hierarchical intelligence
that selects local or remote reasoning according to task feasibility while
preserving a validated fast path for time-critical decisions
\cite{dong2026aerial,nguyen2026agentic}.

\subsection*{B. Scalable Orchestration}

Capability feasibility changes with platform state and network conditions. As
the system grows, no agent can maintain a complete current view. Frequent
updates consume mission resources, whereas a stale view can bind a workflow to
an invalid provider. Concurrent workflows also compete for capabilities, so a
locally attractive binding may reduce the chance that another task can finish.
Hierarchical capability views and task-aware orchestration offer a possible
direction \cite{chen2025ioa,guo2025discovery}. The central question is how much shared state is sufficient without
restoring dependence on global synchronization.

\subsection*{C. Trustworthy Execution}

The validated workflow space itself must be constructed, versioned, and
extended without admitting unsafe or physically infeasible execution paths
\cite{qiao2025worfbench,wang2025awm,li2026graphflow}.
Heterogeneous agents may also interpret the same contract differently, and
their confidence estimates may not be comparable. Evidence can become
outdated or conflict with another observation; it can also be manipulated. A
repaired workflow may appear semantically plausible while violating execution
constraints. Once incorrect evidence activates a transition, the error can
propagate and make responsibility difficult to trace. Progress requires common
machine-readable semantics and runtime verification, while independent safety
rules and human authority remain necessary for high-risk actions.

\subsection*{D. Mobility Co-Design}

UAV movement changes both communication and task observation, so repositioning
cannot be treated as link optimization alone \cite{zeng2016wireless,mozaffari2019tutorial}. A movement may restore a path but
cause the UAV to lose its target, or it may benefit one workflow while
disrupting another. Physical motion also takes longer than network adaptation,
and uncertain prediction can cause unnecessary repositioning. A task-level
co-design should coordinate fast communication changes with slower physical
actions and evaluate whether movement produces a net gain in workflow
continuity.


\section*{\large{VII. Conclusion}}

This article proposed low-altitude agentic networking as a task-level
networking paradigm for organizing distributed UAV intelligence and
capabilities around mission workflows. The architecture separates task logic
from executor identity, derives capability requirements from validated
workflows, and dynamically binds them to agents that remain feasible under
current conditions. The case study reveals a bounded value proposition:
conventional organization remains effective for simple workflows under stable
conditions, whereas dynamic agentic organization becomes beneficial when
workflow dependencies are coupled, capability feasibility changes, and viable
recovery paths remain available. Low-altitude agentic networking therefore
extends networking beyond UAV connectivity and data delivery toward
maintaining the task-specific relations needed to organize distributed
intelligence in dynamic low-altitude environments.




\bibliographystyle{IEEEtran}
\bibliography{bibtex,ref}

\end{document}